# SPATIAL CORRELATION OF LINEAR AND NONLINEAR ELECTRON TRANSPORT IN A SUPERCONDUCTING MICROWAVE RESONATOR: LASER SCANNING MICROSCOPY ANALYSIS

Alexander P. Zhuravel, Steven M. Anlage[1], Stephen Remillard[2], Alexey V. Ustinov[3]

B. Verkin Institute for Low temperature Physics & Engineering, NAS of Ukraine
Address: 47 Lenin Avenue, Kharkov, 61103, Ukraine
Tel.: +380-572-308507, Fax: +380-572-322370, E-mail: zhuravel@ilt.kharkov.ua
[1]Physics Department, Center for Superconductivity Research, University of Maryland
College Park, MD 20742-4111 USA
Tel.: (301) 405-7321, Fax: (301) 405-3779
[2]Agile Devices, 906 University Place, Evanston, IL 60201-3121 USA
Tel.: 847-570-4392, Fax: 847-866-1808, E-mail: remillard@agile-devices.com
[3]Physics Institute III, University of Erlangen-Nuremberg
Erwin-Rommel Str. 1, D-91058, Erlangen, Germany
Tel.: +49 9131 85 27268, Fax: +49 9131 15249, E-mail: ustinov@physik.uni-erlangen.de

Nonlinear (NL) microwave (RF) response of high-$T_C$ superconductors (HTS) at high circulating power limits the applicability of HTS resonators to the few-mW level. *Globally*, the NL response is the integrated contribution of individual *local* sources that are non-uniformly distributed in superconducting elements. Therefore, methods of spatially-resolved RF analysis are needed to identify the NL sources as well as to establish their relative influence on the global response.

Laser scanning microscopy (LSM) has made a good showing as a method for 2-D probing simultaneously the spatial variations of *(i)* RF current flow, $J_{RF}(x,y)$, of *(ii)* areas of resistive dissipation, and *(iii)* the sources of microwave NL in operating HTS devices [1]. A brightness contrast in the LSM images is created by photoresponse (PR) signals arising from local overheating of the HTS film with a micron-size laser probe. The intensity of the probe is modulated by a 100 kHz oscillator producing a periodic modification of the shape of the temperature-dependent microwave transmission $S_{21}(f,T)$ characteristic. At fixed RF frequency, f, and spatially constant temperature oscillation ($\delta T(x,y) = 0.1$ K) in the x,y scanned area of the HTS film, the LSM PR is proportional to the laser-beam-induced changes in resonator transmittance, $\delta //S_{21}(f)//^2$, that can be expressed in a form [2]:

$$PR \sim \delta \|S_{12}(f)\|^2 = \frac{1}{2}\left( \frac{\|S_{12}(f)\|^2}{\partial f_0} \frac{\partial f_0}{\partial T} + \frac{\|S_{12}(f)\|^2}{\partial (1/2Q)} \frac{\partial (1/2Q)}{\partial T} + \frac{\|S_{12}(f)\|^2}{\partial \hat{S}_{12}^2} \frac{\partial \hat{S}_{12}^2}{\partial T} \right) \delta T(x,y), \qquad (1)$$

where the three items in the brackets in (1) symbolize inductive ($PR_X$), resistive ($PR_R$), and insertion loss ($PR_{IL}$) components of LSM PR, respectively, and $f_0$ – is the resonant RF frequency, Q – is the quality factor, and $\hat{S}_{12}$ – is the maximum of the transmission coefficient. Evidently, the $PR_X(x,y)$ originates from frequency $\delta f_0$ tuning due to the laser-probe-induced modulation of the HTS kinetic inductance. The effect was used to image a quantity proportional to the square of the local current density $J_{RF}(x,y)$. The remaining components $PR_R(x,y)$ and $PR_{IL}(x,y)$ reflect changes in Ohmic dissipation produced by the laser probe and, therefore, were used to image spatial variations of the dissipation by a procedure that is described in detail in [2].

The two-tone method of excitation by two microwave sources having fixed frequencies $f_{1,2} = f_0 \pm \Delta f/2$ and the same output power $P_{f1}=P_{f2}$ was applied to image RF properties of the resonator, where $f_0 = 1.872$ GHz is the frequency of the fundamental resonance at T = 77 K, and $\Delta f = 0.2$ MHz is the spacing. The zero-span mode of a Spectrum Analyzer was used to detect the probe-induced LSM PR(x,y) at $f_1$ and $f_2$ for imaging $PR_R(x,y)$ and $PR_{IL}(x,y)$ in compliance with the partition method described in [2]. The modulation of RF signals transmitted at $f=2f_1-f_2$ and at $f=2f_2-f_1$ intermodulation (IMD) harmonics were used to image the distribution of *microscopic* sources of NL response [3].

Figure 1(a) shows 2-D LSM $PR_X(x,y)$ image of the HTS device previously characterized by us in [3]. This was a $YBa_2Cu_3O_y$ (YBCO) film with thickness of about 1 $\mu$m configured by ion-milling lithography on $LaAlO_3$ (LAO) substrate into a meandering strip line resonator with line width of 250 $\mu$m. The topology of the resonator is clear from the $PR_X(x,y)$ map manifesting the predicted peaks of $J_{RF}(x,y)$ along the strip edges. Ends of the micro strip are outlined by dotted line in sections A and C in the Fig. 1(a) for clarity. At P = 10 dBm, the $PR_X(x,y)$ reached a maximum corresponding to $J_{RF}(x,y) = 6.35*10^5$ A/cm$^2$ at the position of a corner outlined by the white box D. To be sure that the imaged $J_{RF}(x,y)$ in Fig. 1(a) corresponds exactly to the distribution of RF electronic transport in the resonator standing wave pattern, an unfolding of the section-averaged $(PR_X)^{1/2}$ distribution is plotted in Fig. 1(b). As evident, a fit to $J_{RF}(L)= J_{RF}^{MAX}\sin(2\pi L/L_0)$ ($L_0$ = length of meandering segment from A to C) is almost ideally matched to the measured distribution along the length L of the resonator from A through mid section B to C, showing averaged $J_{RF}(L=B)=1.3*10^5$ A/cm$^2$ at the crest of standing wave. This value was used to calibrate the amplitude of $PR_X$ in $J_{RF}$ units. A line-scan profile of the $PR_X(x)$ distribution in section B was used to make such a calibration [1].

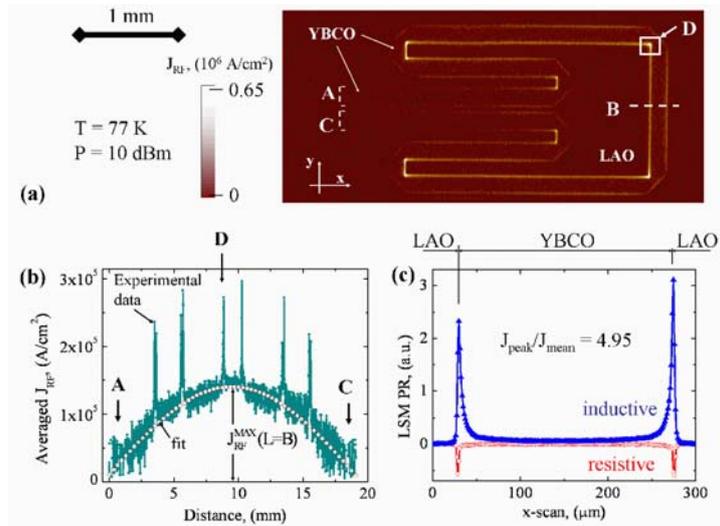

Fig.1. (a) 2-D LSM map of $J_{RF}(x,y)$, (b) standing wave pattern of section-averaged RF current density along the length of the strip line, and (c) mid-section profiles of $PR_X(x)$ and $PR_R(x)$ at T = 77K, P=10 dBm, and a loaded $Q_L \sim 1860$. Brighter areas in (a) correspond to peak values of $J_{RF}(x,y)$

Figure 2(a) shows RF power-dependent modification of the edge profiles of the total LSM PR. A series of profiles was obtained at $f_1$ (positive PR amplitude, above $f_0$) and at $f_2$ (negative PR amplitude, below $f_0$) by repeating the same line scan through the corner D in the range of input P from -12dBm to +15 dBm in 1 dBm steps. It was detected that in linear LSM PR mode the photoresponse grows in a nonlinear manner at high P starting from 4 mW (~2 dBm). Moreover, the shape of the $PR(x, f_1)$ profile was spatially collapsed at P>8 dBm (see the 10 dBm profile in Fig. 2(a)). This is due to the impact of the negative sign of $PR_R(x)$ at all the frequencies decreasing the amplitude of the total PR in resistive areas at $f_1$ and increasing it at $f_2$. By focusing the laser probe at an arbitrary position on the HTS film in the vicinity of D, we extracted the local values of different components of LSM PR in an area limited by the radius of the thermal healing length ($l_T \sim 4$ $\mu$m) around the best focus. The normalized power dependencies of those $PR_X$, $PR_R$, and $PR_{IMD}$ are shown in Fig.2 (b). One can see an almost linear growth of $PR_X$ as a function of increasing input power. In contrast, $PR_R$ remains zero up to some critical power $P_C \sim 2$ mW corresponding to $J_C(D)=1.02$ MA/cm$^2$ and then starts its NL behavior for $J_{RF}(D)>J_C(D)$. This feature correlates well with the power dependence of NL ($PR_{IMD}$) components, perhaps showing an interconnection of $J_C(D)$ and the NL current density scale, $J_{NL}(D)$. However, the onset of $PR_R(P)$ here produces a maximum in the $PR_{IMD}(P)$ dependence that may result from a decreasing of the relative portion of superfluid component in the total $J_{RF}$ due to depairing, with $J_{NL} \sim 1$ MA/cm$^2$ at the corner D. The effect of Cooper depairing is spatially dependent since it is directly determined by both the topology of $J_{RF}(x,y)$ and the superconducting gap distribution at defects of the HTS microstructure and patterned edges, as well as local doping.

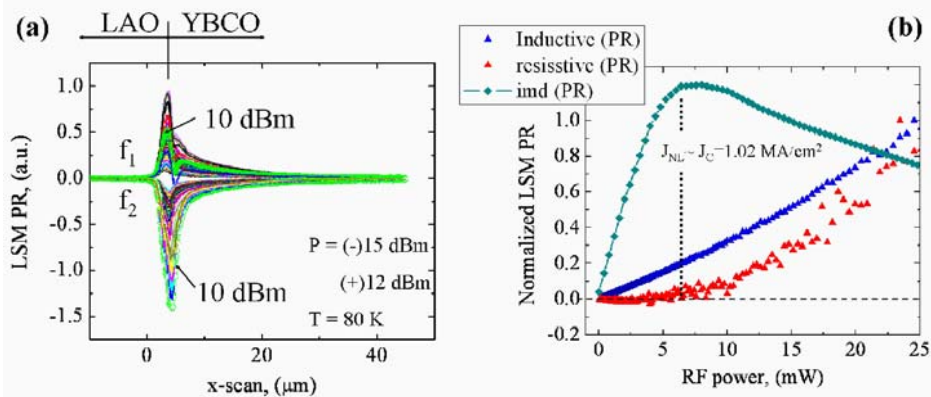

Fig. 2. (a) Detailed profiles of LSM PR(x) through area D obtained at frequencies $f_1$ and $f_2$ in the range from -12dBm to +15dBm, and (b) extracted local power dependencies of $PR_X$, $PR_R$, and $PR_{IMD}$

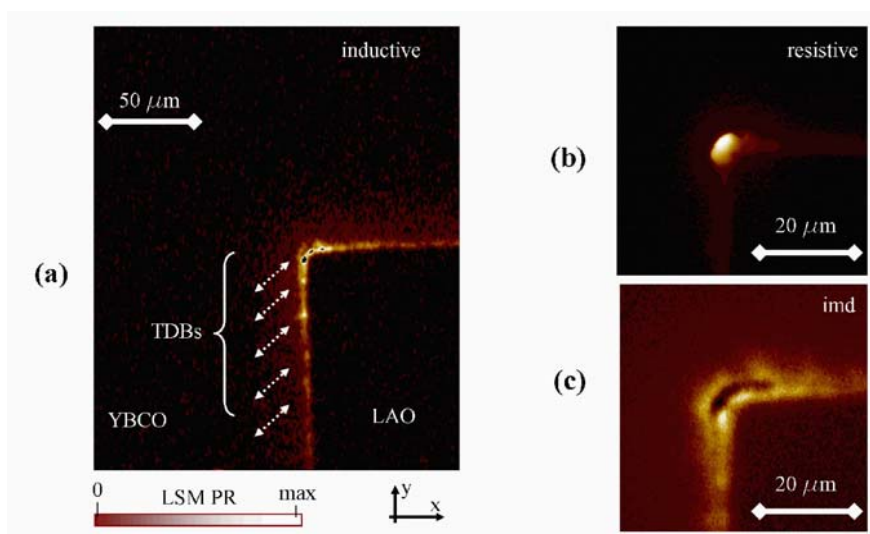

Fig. 3. (a) 2-D LSM images of (a) $PR_X$, (b) $PR_R$, and (c) $PR_{IMD}$ in the vicinity of area D in Fig. 1(a)

Figure 3(a) shows a 200 x 250 μm$^2$ area scan of $PR_X(x,y)$ in region D. Spatial modulation of $PR_X(x,y)$ is obvious here owing to the highest $J_{RF}(x,y)$ along the microstrip edges and due to variations of magnetic penetration depth at low-angle grain-boundaries (GBs) formed at interfaces of individual twin-domain blocks (TDBs). The direction and position of some TDBs are marked by double-arrow lines in the Fig. 3(a). Those TDBs (retracing the surface structure of the LAO substrate) were identified independently by the LSM operating in the reflection mode of a light microscope. Additionally, a detailed $PR_R(x,y)$ image is shown in Fig. 3(b). The distinguishing feature of this image is that the edge inhomogeneities of $PR_R(x,y)$ are obscured by the very large LSM PR just at the corner D, resulting from exponential growth of local current-voltage steepness from $J_{RF}(x,y)$. In this area the maximum of $J_{RF}(x,y)$ may be creating an overcritical state of the HTS film leading to a more linear (normal-metal like) behavior of the local NL sources that is seen as a black (zero-response) zone in the IMD image [see Fig. 3(c)]. In contrast, the remaining critical-state regions give a maximum value to the $J_{IMD}(x,y)$ signal as evident from Fig. 3(c).

**References**

1. A.P. Zhuravel, A.G. Sivakov, O.G. Turutanov, A.N. Omelyanchouk, Steven M. Anlage, A. Lukashenko, A.V. Ustinov, D. Abraimov, "Laser scanning microscopy of HTS films and devices (Review Article)", Low Temperature Physics **32**, pp. 592-607 (2006);
2. A.P. Zhuravel, Steven M. Anlage, A. V. Ustinov, "Measurement of local reactive and resistive photoresponse of a superconducting microwave device", Appl. Phys. Lett. **88**, p. 212503 (2006);
3. A.P. Zhuravel, S.K. Remillard, Steven M. Anlage, and A.V. Ustinov, "Spatially Resolved Analyses of Microwave and Intermodulation Current Flow across HTS Resonator Using Low Temperature Laser Scanning Microscopy", in Book:"MSMW'04 Symposium Proceedings. Kharkov, Ukraine", **V.1**, p.p. 421-423 (2004).